\documentclass[aps,prb,showpacs,twocolumn]{revtex4}

\usepackage{graphicx}

\usepackage{amsmath} \newcommand{\EqLabel}[1]{\label{#1}}

\begin{document}
\title{Heat transport in quantum spin chains: the relevance of
integrability}

\author{Jinshan Wu and Mona Berciu}

\affiliation{ Department of Physics and Astronomy, University of
  British Columbia, Vancouver, BC, Canada, V6T~1Z1}

\begin{abstract}
We investigate heat transport in various quantum spin chains, using the
projector operator technique. We find that anomalous heat transport 
is linked not to the integrability of the Hamiltonian, but to whether
it can be mapped to a model of non-interacting fermions. Our results also suggest
how seemingly anomalous transport may occur at low temperatures in
a much wider class of  models.
\end{abstract}

 \date{\today}

\maketitle

Heat transport in quantum spin chains, in particular when is normal
(diffusive) transport observed, is still not understood despite
considerable effort.\cite{Review2007,Review_Drude,Heidrich_XXZ,
  Klumper_XXZ, Shastry_OBC, Saito_Kubo, Heidrich_XXZB,Zotos_Drude,
  Shastry, Ian, Tomaz} For example, it was conjectured that
integrability leads to anomalous (ballistic)
transport,\cite{Zotos_Drude} but it was also argued that an integrable
gaped XXZ chain has normal conductivity.\cite{Shastry} Others have
argued that only the spin conductivity is normal in this case, while
the thermal conductivity is still anomalous.\cite{Tomaz} A consensus
on what are the necessary criteria for normal conductivity is still
missing.

Most of the above work\cite{Heidrich_XXZ, Klumper_XXZ, Shastry_OBC,
  Saito_Kubo, Heidrich_XXZB,Zotos_Drude, Shastry, Ian} studied
infinite and/or periodic chains, and used the Kubo formula\cite{Kubo,
  Luttinger} where finite/zero Drude weight signals anomalous/normal
transport. For integrable systems, the Kubo
formula always predicts anomalous heat transport. In fact, full
integrability is not even necessary, all that is needed is commutation
of the heat current operator with the total
Hamiltonian.\cite{Klumper_XXZ} Anomalous heat transport observed
experimentally in systems described by integrable models, such as
(Sr,Ca)$_{14}$Cu$_{24}$O$_{41}$, Sr$_2$CuO$_3$ and
CuGeO$_3$,\cite{e1,e2,e3} seems to validate this result, although
Ref. \onlinecite{e4} finds normal transport in Sr$_2$CuO$_3$ at high temperatures.

The proof of the Kubo formula requires dealing with currents between
the chain's ends and the thermal baths it is connected to.\cite{notem}
For infinite systems, one may argue that such currents can be ignored,
as they are a boundary effect. However, the terms describing the
coupling to the baths lead to a non-vanishing commutator between the
heat current operator and the total Hamiltonian, invalidating the main
argument for anomalous transport. In other words, ``integrability'' of
the chain connected to baths may be lost even if the isolated chain is
integrable. This conclusion is supported by recent proofs of Kubo-type
formulae for finite systems, based on phenomenological approaches such
as the Fokker-Planck equation.\cite{openKubo}  These Kubo formulae
have similar structure to the original one, however the dynamics is
not defined only by Hamiltonian of the chain but also includes random
variables mimicking the effects of coupling to the baths.

Here we investigate finite spin chains coupled to thermal baths, using
the projector technique.\cite{KuboBook, Saito_Projector, Li,
  Mahler_Projector} There are many other similar
studies\cite{Saito_Projector, Li, Mahler_Projector, MM, Saito_MD_EPL,
  Mahler_Local, Michel_HAM, Tomaz} for various spin Hamiltonians, most
of which would be integrable for a periodic, isolated chain (hereafter
we call such models integrable).  Results range from normal to
anomalous transport, and there is no agreement on whether
integrability is correlated or not with anomalous transport.

We propose a resolution for this question in this Rapid
Communication. We find that integrability is not a sufficient
condition for anomalous heat transport. We find anomalous transport at
all temperatures only in models which can be mapped onto homogeneous
non-interacting fermionic models. All other models we investigated
exhibit normal heat transport, whether they are integrable or not
(however, as discussed below, at low temperatures their heat transport
may become anomalous in certain conditions). We therefore conjecture
that the existence of such a mapping is the criterion determining
anomalous transport, at least for finite-size systems.

We begin by briefly describing our calculation method, which is a
direct generalization of the projection operator technique
used to study the evolution towards equilibrium of a system coupled
to a single bath.\cite{KuboBook} The $N$-site chain of spins-${1\over
2}$ is described by the Hamiltonian:
\begin{equation}
{\cal H}_S = \sum_{i=1}^{N-1} \left[J_{x}s_{i}^{x}s_{i+1}^{x} +
J_{y}s_{i}^{y}s_{i+1}^{y} + J_{z}s_{i}^{z}s_{i+1}^{z}\right] - \vec{B}
\sum_{i=1}^{N}\vec{s}_{i}
\nonumber \label{spinhamiltonian}
\end{equation}
while the heat baths are collections of bosonic modes:
\begin{align}
\nonumber 
{\cal H}_B = \sum_{k, \alpha} \omega_{k,\alpha}
b^{\dag}_{k,\alpha}b_{k,\alpha}
\end{align}
where $\alpha=R/L$ indexes the right/left-side baths and we set
$\hbar=1, k_B=1$, and the lattice constant $a=1$.  The system-baths
coupling is taken as:
\begin{equation}
\nonumber 
V = \lambda \sum_{k,\alpha} V^{(\alpha)}_{k}s_{i_\alpha}^{y}\otimes
\left(b^{\dag}_{k,\alpha} + b_{k,\alpha}\right)
\end{equation}
where $i_L =1$ and $i_R=N$, {\em i.e.} the left (right) thermal bath is only
coupled to the first (last) spin and can induce its
spin-flipping. This is because we choose $\vec{B}\cdot \vec{e}_y =0$
while $|\vec{B}|$ is finite, meaning that spins primarily lie in the
$x0z$ plane so that $s^{y}$ acts as a spin-flip operator.

The evolution of the total system is described by the Liouville-von
Neumann equation for the total density matrix $\hat{\rho}_T$.  If we
are interested only in the properties of the central system, it is
convenient to find an equation of motion for the reduced density
matrix $\hat{\rho}_{S}=tr^{B}\left(\hat{\rho}_T\right)$ and solve it
directly. This is achieved by using the projection operators, treating
the system-bath coupling perturbationally to second order, and also by
using a Markovian approximation.\cite{KuboBook, Saito_Projector} These
approximations are reasonable: the system-baths coupling must be
rather weak so that the properties of the chain are determined by its
specific Hamiltonian, not by this coupling. The Markovian
approximation is also justified, since we are interested only in the
steady-state limit $t \rightarrow \infty$.

The resulting equation of motion for $\hat{\rho}_{S}\left(t\right)$ is:
\begin{equation}
\EqLabel{evol} \frac{\partial \hat{\rho}_S(t)}{\partial t} =-i[{\cal
H}_S, \hat{\rho}_S(t)]-\lambda^2\sum_{\alpha=L,R}^{}\left(
\left[s_{i_\alpha}^{y}, \hat{m}_{\alpha} \hat{\rho}_S(t)\right] +
h.c.\right)
\end{equation}
where $ \hat{m}_\alpha = s_{i_\alpha}^{y} \cdot
\hat{\Sigma}_{\alpha} $.  Here, $(\cdot )$ refers to the element-wise
product of two matrices, $\langle n| \hat{a}\cdot \hat{b} |m\rangle =
\langle n| \hat{a} |m\rangle \langle n|\hat{b} |m\rangle$. The bath
matrices $\hat{\Sigma}_{L,R}$ are defined in terms of the eigenstates
of the system's Hamiltonian ${\cal H}_S|n\rangle = E_n |n\rangle$ as:
\begin{eqnarray}
\nonumber \hat{\Sigma}_\alpha = \pi \sum_{m,n}^{} |m\rangle\langle n|
\left[\Theta\left(\Omega_{mn}\right)n_\alpha\left(\Omega_{mn}\right)
D_\alpha\left(\Omega_{mn}\right)|V^{(\alpha)}_{k_{mn}}|^2 \right. &&\\
\nonumber \left.  + \Theta\left(\Omega_{nm}\right)\left( 1+
n_\alpha\left(\Omega_{nm}\right)\right)
D_\alpha\left(\Omega_{nm}\right)|V^{(\alpha)}_{k_{nm}}|^2 \right] &&
\end{eqnarray}
where $\Omega_{mn} = E_{m}-E_{n}=-\Omega_{nm} $ and $k_{mn}$ is
defined by $\omega_{k_{mn},\alpha} = \Omega_{mn}$, {\em i.e.} a bath mode
resonant with this transition. Furthermore, $\Theta(x)$ is the
Heaviside function, $n_\alpha(\Omega)=\left[e^{\beta_\alpha \Omega}
-1\right]^{-1}$ is the Bose-Einstein equilibrium distribution for the
bosonic modes of energy $\Omega$ at the bath temperature $T_\alpha
=1/\beta_\alpha$, and $D_\alpha(\Omega)$ is the bath's density of
states. The product
$D_\alpha\left(\Omega_{mn}\right)|V^{(\alpha)}_{k_{mn}}|^2$ is the
bath's spectral density function. For simplicity, we take it to be
a constant independent of $m$ and $n$. 

The same equation of motion for $\hat{\rho}_S(t)$ was also derived in
Refs.~\onlinecite{Saito_Projector, Mahler_Projector, Li}, where its
steady-state solution was found via 
Runge-Kutta integration or by solving an eigenvalue problem. The latter
comes about because the steady-state $\hat{\rho}_{\infty}$  is given by 
$0={\cal L} \hat{\rho}_{\infty}$, where ${\cal L}$ is the linear
operator on the RHS of Eq. (\ref{evol}), so in matrix terms
$\rho_\infty$ is the eigenvector corresponding to the zero eigenvalue
of  ${\cal L}$. Using the normalization  $tr
\hat{\rho}_{\infty}=1$, this eigenvalue problem can be replaced with
solving a linear system of coupled equations, which makes it more
efficient and allows us to analyze somewhat larger systems.

We rewrite
${\cal H}_S = \sum_{i=1}^{N-1} h_{i,i+1} + \sum_{i=1}^{N} h_{i}$,
where $h_{i,i+1}$ is the exchange between nearest-neighbor spins and
$h_{i}$ is the on-site coupling to the 
magnetic field. We can then define a local site Hamiltonian
$h^{(S)}_i = {1\over 2} h_{i-1,i} + h_{i} + {1\over 2} h_{i,i+1}$
(with $h_{0,1}=h_{N,N+1}=0$) and a local bond Hamiltonian
$h^{(B)}_{i}= {1\over 2} h_i + h_{i,i+1} +{1\over 2} h_{i+1}$ such
that ${\cal H}_S =\sum_{i=1}^{N} h^{(S)}_i = \sum_{i=1}^{N-1}
h^{(B)}_{i}$. The local site Hamiltonians can be used to derive
the heat current operator from the continuity equation
$\hat{j}_{i\rightarrow i+1}-\hat{j}_{i-1\rightarrow i} = \nabla
\hat{j} = -\frac{\partial h^{(B)}_i}{\partial t}=-i \left[{\cal
H}_S,h^{(B)}_{i}\right]$. This results in 
$\hat{j}_{i\rightarrow i+1}=i\left[h^{(B)}_{i}, h^{(B)}_{i+1}\right]$
for  $i=1,\cdots, N-2$. 
As expected, in the steady state  we find 
$\langle\hat{j}_{i\rightarrow i+1}\rangle=  J$ to be independent of $i$.

Knowledge of the steady state heat current $J$, as such, is not enough
to decide whether the transport is normal or not.  Consider an analogy
with charge transport in a metal connected to two biased leads. What
shows if the transport is anomalous is the profile of the electric
potential, not the value of the electric current. In anomalous
transport (clean, non-interacting metal) all the voltage drop occurs
at the ends of the sample, near the contacts. Away from these contact
regions, electrons move ballistically and the electric potential is
constant, implying zero intrinsic resistance. For a dirty metal,
scattering takes place everywhere inside the sample and the electric
potential decreases monotonically in between the contact regions, {\em
  i.e.}  the sample has finite intrinsic resistivity.

In principle, the scaling of the current with the sample size, for a
fixed effective bias, also reveals the type of transport: for anomalous
transport, the current is independent of the sample size once its
length exceeds the sum of the two contact regions, while for normal
transport it decreases like inverse length. The problem is that
one needs to fix the effective bias, {\em i.e.}  the difference between
the applied bias and that in the contact regions. Furthermore, since we
can only study relatively short 
chains, the results of such scaling may be questionable.

It is therefore desirable to use the equivalent of the electric
potential for heat transport and to calculate its profile along in
order to determine the type of transport. This, of course, is the
``local temperature'', which is a difficult quantity to define. One
consistency condition for any definition is that if
$T_L=T_R=T$, {\em i.e.} the system is in thermal equilibrium at $T$,
then all local temperatures should equal $T$. We
define local site temperatures $T_i$ which fulfill this
condition in the following way. Since we know all
eigenstates of ${\cal H}_S$, it is straightforward to calculate its
equilibrium density matrix at a given $T$, $\hat{\rho}_S^{eq,T} =
{1\over Z} \sum_{n}^{} e^{-\beta E_n} |n\rangle \langle n|$, where
$Z=\sum_{n}^{} e^{-\beta E_n}$. Let then $\langle h_i^{(S)}\rangle_{eq, T}
= tr [\hat{\rho}_S^{eq,T} h_i^{(S)}]$. We define $T_i$ to
be the solution of the equation: $\langle h_i^{(S)}\rangle_{eq, T_i}=
tr[\hat{\rho}_\infty h_i^{(S)}]$. In other words, 
 the steady-state value of the energy at that site equals the
energy the site would have if the whole system was in equilibrium at
$T_i$. Of course, we can also use other ``local'' operators such as
$h_{i}^{(B)}$ to calculate a local  
bond temperature $T_{i+{1\over 2}}$. We find that when these definitions
are meaningful, the results are in very good agreement no matter what
``local'' operator is used.

This type of definition of $T_i$ is meaningful only if a
large magnetic field $B$ is applied. For small $B$, the
expectation values $\langle h_i^{(S)}\rangle_{eq, T}$ are very
weakly $T$-dependent, so that tiny 
numerical errors in the steady-state value can lead
to huge variations in $T_i$. Addition of a large $B$ is needed to
obtain $\langle h_i^{(S)}\rangle_{eq, T}$ which varies fast enough
with $T$ for values of interest so that a meaningful $T_i$ can be
extracted. Since we could not find a meaningful definition for $T_i$
when $\vec{B}\rightarrow0$, we cannot investigate such cases. Note, however,
that most integrable models remain integrable under addition of an
external field $\vec{B}=B
\hat{e}_z$.

In all of our calculations, we take $B_{z}=1$ and the exchange
$J\sim 0.1$.  Temperatures $T_{L/R}=T(1\pm \delta /2) $ should not be so
large that the steady state is insensitive to the model or so small that
only the ground-state is activated. Reasonable choices lie between
$\min(J_x,J_y,J_z)$ and $NB$, which are roughly the
smallest, respectively the largest energy scales for an $N$-site spin
chain. 

\begin{figure}[t]
\includegraphics[width=0.85\columnwidth]{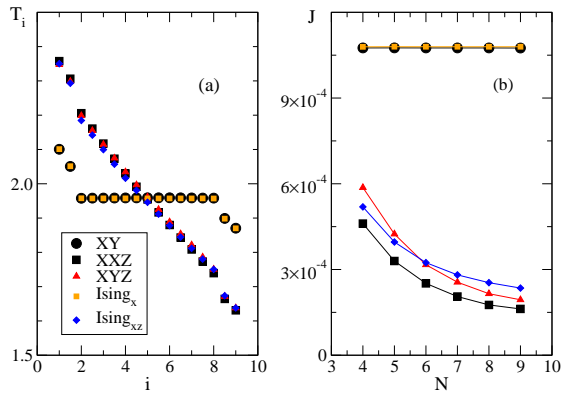}
\caption{(color online) Plot of (a) local temperature profile 
for   chains with $N=9$, and (b) thermal current $J$ vs. chain
length $N$. In all cases $T_L=2.4, T_R=1.6$. Other non-zero
parameters are $J_x=J_y=0.1, B_z=1$ (XY); $J_x=J_y={1\over 2} J_z=0.1, 
B_z=1$ (XXZ); $J_x={1\over 2}J_y={1\over 3} J_z=0.1,
B_z=1$ (XYZ); $J_z=0.2, B_x=1$ (Ising$_x$) and 
$J_z=0.2, B_x=B_z=1$ (Ising$_{xz}$). The  $XY$ 
and the  Ising$_{x}$ chains show flat $T_i$ profiles and 
currents independent of the chain's length, {\em i.e.} anomalous
transport. All other models have normal transport. 
} 
\label{fig1} 
\end{figure}

In Fig. \ref{fig1}(a) we show typical results for local temperature
profiles $T_i$, $T_{i+{1\over 2}}$. We apply a large bias
$\delta=(T_L-T_R)/T=0.4$ for clarity, but we find similar results for
smaller $\delta$ (see below). For these values,  the
``contact regions'' include about 
two spins on either end. The $T_i$ profile of the rest of the chain is
consistent with anomalous transport (flat $T_i$ profile) for the $XY$
chain and shows normal transport (roughly linear
$T_i$ profile) in all the other non-Ising, $J_z\ne 0$  cases. We find similar
results (not shown) for ferromagnetic couplings. All these are
integrable models. The $XY$ model is special because it can be
mapped to non-interacting spinless fermions with the
Jordan-Wigner transformation.\cite{JW1} A finite $J_z$ leads to
nearest-neighbor interactions between  fermions. 
Eigenmodes for $J_z\ne 0$ can  be found using Bethe's
ansatz, but they cannot be mapped to non-interacting fermions.

Another model that maps to non-interacting spinless fermions is the
Ising model in a transverse field $B_x$.\cite{JW2} For this
model we again find anomalous transport, as shown in Fig.
\ref{fig1}(a). If we add a $B_z$ field, the model becomes
non-integrable\cite{comm2} and we recover normal transport. The scaling
of  $J$ vs. $N$, shown in Fig.
\ref{fig1}(b), supports these conclusions, although a quantitative analysis
is difficult because of the  contact regions's contributions.

We found this generic behavior for a wide range of parameters. When
$\lambda\in\left[0.03, 0.2\right]$, $T\in\left[0.3, 30.0\right]$ and
$\delta  \geq 0.01$, the XYZ model has normal conductivity when
$J_{z} \in \left[0.03, 0.5\right]$ and anomalous conductivity when
$J_{z}=0$. When $0< J_{z}<0.03$ or $J_z \in [0.5, 1]$ the local
temperature still decreases monotonically but not linearly, and the
current decreases more slowly than $1/N$. Since we cannot study much longer
chains we cannot easily distinguish here between normal vs. anomalous
transport. For $J_z \ge 1$ the system becomes
Ising-like and the transport is anomalous, as expected.

We would like to better gauge how things change with 
$N$, even with  our limited
range. For this, we consider how the effective
temperature bias    on the chain, $T_{2}-T_{N-1}$, or its effective slope
$S=\frac{T_{2}-T_{N-1}}{N-3}$, depend on the system size $N$.  For
 normal transport we expect $S \propto
(N-\beta)^{-1}$ ($\beta$ accounts for the contact regions). If $S
 \propto (N-\beta)^{-\alpha}$ with $\alpha >1$, then for 
longer chains the temperature profile tends to be flatter than normal so we will
use $\alpha >1$ as a signature of anomalous transport (of course, if
a plateau starts to emerge near the center of chain that also indicates
anomalous transport). A second gauge of the size dependence comes
from looking at how the shape of the normalized temperature profile changes with
$N$.

\begin{figure}[t]
\includegraphics[width=0.85\columnwidth]{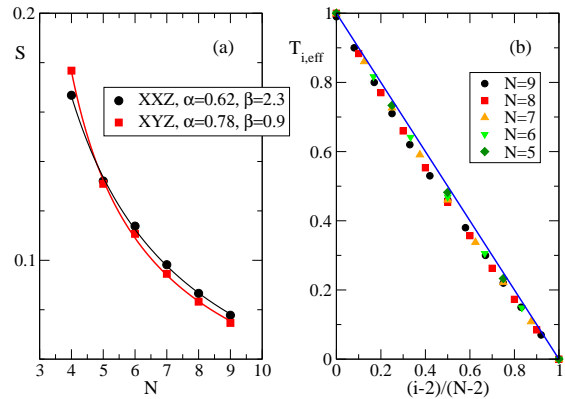}
\caption{(a) Slope $S$ vs. system size $N$, and (b) normalized
  temperature $T_{i,{\rm eff}}=(T_i-T_{N-1})/(T_2-T_{N-1})$ for various systems sizes $N$. Parameters are
  $J_{x}=J_{y}=0.1$, $J_{z}=0.2$, $B_{z}=1.0$, $T=2.0$, $\delta =0.2$,
  $\lambda=0.1$. See text for more details.} \label{fig2} 
\end{figure}

Figures \ref{fig2}(a) and (b) show such analysis. The left panel
shows fits for $S$ (solid lines) for XXZ and  XYZ models
(symbols). Best fits give $\alpha <1$, consistent with normal 
transport. Similarly, the right panel which plots
the normalized temperature profiles for different values of $N$ shows
no change with increasing $N$, and  no evidence
that a plateau may evolve. Based on this limited evidence,
we conclude that these models, although integrable, do exhibit
normal transport. 
  
In summary, the first conclusion we draw from these results is
that integrability is not sufficient to guarantee anomalous transport:
several integrable models show normal heat transport, in agreement
with other studies.\cite{MM, Mahler_Projector, Michel_HAM, Ian} The
second conclusion is that only models that map onto Hamiltonians of
non-interacting fermions exhibit anomalous heat transport. This is a
reasonable sufficient condition, since once inside the sample (past
the contact regions) such fermions propagate ballistically. However,
we cannot, at this stage, demonstrate that this is a necessary condition
as well. We therefore can only conjecture that this is the criterion
determining whether the heat transport is anomalous.

In this context, it is important to emphasize again the essential role
played by the connection to the baths. In its absence, an isolated
integrable model is described by Bethe ansatz type
wavefunctions. Diffusion is impossible since the conservation of
momentum and energy guarantees that, upon scattering, pairs of
fermions either keep or interchange their momenta.  For a system
connected to baths, however, fermions are continuously exchanged with
the baths, and the survival of a Bethe ansatz type of wavefunction
becomes impossible. In fact, even the total momentum is no longer a
good quantum number. We believe that this explains why normal transport
in systems mapping to interacting fermions is plausible.

Normal transport is also possible for non-interacting fermions, if
they are subject to elastic scattering on disorder.  This can 
be realized, for example, by adding to the $XY$ model a random field
$B_z$ at various sites. We have verified (not shown) that a local drop in the
local temperature indeed arises near sites with such disorder, leading
to normal conductance in ``dirty'' samples.

On the other hand, anomalous transport can also occur in models which
map to homogeneous interacting fermions if the bath temperatures are
very low. Specifically, consider the XXZ models. Because of the large
$B_z$ we use, the ground-state of the isolated chain is ferromagnetic
with all spins up. The first manifold of low-energy eigenstates have
one spin flipped (single magnon states), followed by states with two
spins flipped (two magnon states), etc. The separation between these
manifolds is roughly $B_z$, although because of the exchange terms
each manifold has a fairly considerable spread in energies and usually
overlaps partially with other manifolds. 

If both $T_L, T_R \ll B_z$,
only single-magnon states participate in the
transport. We can then study numerically very long chains by assuming
that the steady-state matrix elements $\rho_{nm}$ vanish for all other
eigenstates ($S_{z,tot}$ is a good quantum number for these
models). In this case we find anomalous transport for all 
models, whether integrable or not. This is reasonable, since the lone
magnon (fermion) injected 
on the chain has nothing else to interact with, so it 
must propagate ballistically. 

We can repeat this restricted calculation by including the two-magnon,
three-magnon, etc.  manifolds in the computation. As expected, the
results agree at low $T_L, T_R$, but differences appear for
higher $T_L, T_R$, when these higher-energy manifolds become thermally
activated. In such cases, the
transport becomes normal for the models mapping to interacting
fermions as soon as the probability to be in the two (or more) magnon
sector becomes finite. In other words, as soon as multiple excitations
(fermions) are simultaneously on the chain, and 
inelastic scattering between them becomes possible.

These results may explain the heat transport observed experimentally in compounds
such as Sr$_2$CuO$_3$,\cite{e4} where at low
temperature anomalous transport was found while at high temperature
normal transport was reported.

In conclusion, we propose a new conjecture for what determines the
appearance of anomalous heat transport at all temperatures in spin
chains. Unlike previous suggestions linking it to the
integrability of the Hamiltonian or existence of gaps, we propose that the criterion
is the  mapping of the Hamiltonian onto a model of
non-interacting fermions without any disorder.

{\em Acknowledgments}: Discussions and suggestions from Ian
Affleck are gratefully acknowledged. Work supported by 
 CIfAR Nanoelectronics,
CFI and NSERC.


\begin{thebibliography}{99}
\bibitem{notem} Thermal transport is driven by connecting the ends of
  the system to thermal baths, not by inducing a variation of the gravitational
  potential across the entire chain.
\end{thebibliography}
\end{document}